\documentclass[conference]{IEEEtran}
\ifCLASSINFOpdf
\else
\fi
\usepackage{amsfonts}
\usepackage[dvips]{graphicx}
\usepackage{times}
\usepackage{cite}
\usepackage{amsmath}
\usepackage{array}
\usepackage{amssymb}

\newcommand{\bs}{\boldsymbol}

\usepackage{stfloats}
\usepackage{slashbox}
\usepackage{graphicx}
\usepackage{footnote}
\usepackage{booktabs}
\usepackage{array}
\usepackage{algorithmic}
\usepackage{algorithm}
\usepackage{subeqnarray}
\usepackage{cases}
\usepackage{threeparttable}
\usepackage{color}
\IEEEoverridecommandlockouts

\newtheorem{corollary}{Corollary}
\begin{document}
%
\title{Secrecy Wireless Information and Power Transfer in OFDMA Systems}

\author{\IEEEauthorblockN{Meng Zhang\IEEEauthorrefmark{1},
 Yuan Liu\IEEEauthorrefmark{1},
 and Rui Zhang\IEEEauthorrefmark{2}}
 \vspace*{0.5em}
\IEEEauthorblockA{\IEEEauthorrefmark{1}School of Electronics and Information Engineering,
South China University of Technology, Guangzhou, 510641, P. R. China}
\IEEEauthorblockA{\IEEEauthorrefmark{2}Department of Electrical and Computer Engineering, National University of Singapore, Singapore\\
Email: jackey\_zm@hotmail.com, eeyliu@scut.edu.cn, elezhang@nus.edu.sg}
\thanks
{This work is supported in part by the Natural Science Foundation of China under Grant 61401159.}}


%


\maketitle

\begin{abstract}
 In this paper, we consider simultaneous wireless information and power transfer (SWIPT) in orthogonal frequency division multiple access (OFDMA) systems with the coexistence of information receivers (IRs) and energy receivers (ERs). The IRs are served with best-effort secrecy data and the ERs harvest energy with minimum required harvested power. To enhance physical-layer security and yet satisfy energy harvesting requirements, we introduce a new frequency-domain artificial noise based approach. We study the optimal resource allocation for the weighted sum secrecy rate maximization via transmit power and subcarrier allocation. The considered problem is non-convex, while we propose an efficient algorithm for solving it based on Lagrange duality method. Simulation results illustrate the effectiveness of the proposed algorithm as compared against other heuristic schemes.
\end{abstract}


%
\IEEEpeerreviewmaketitle

\section{Introduction}
Orthogonal frequency division multiple access (OFDMA) gains its popularity due to its flexibility in resource allocation and robustness against multipath fading, and has become a promising multi-access technique for multiuser wireless communications networks.

Recently, simultaneous wireless information and power transfer (SWIPT) has appeared as an appealing solution to prolong the lifetime of energy-constrained wireless nodes, by enabling them to receive energy and information from the same signal. SWIPT has drawn a great deal of research interests \cite{ZhouZhang,DerrickIC13,ZhangHo}. For instance, in \cite{ZhouZhang}, the authors studied two practical schemes for orthogonal frequency division multiplexing (OFDM) based SWIPT, namely, power splitting and time switching. With time switching applied at each receiver, the received signal is processed either for energy harvesting or for information decoding at any time. When power splitting is employed at the receiver, the signal is split into two streams, then processed for energy harvesting and information decoding, respectively. In \cite{DerrickIC13}, the authors considered a multiuser OFDM system with some users to decode information, and the remaining users to harvest energy.

Besides, due to the increasing importance of information security, substantial works have been dedicated to information-theoretic physical layer (PHY) security (e.g. \cite{Shannon,Wang,Renna,MengTIFS,Goel,Qin13}), as a complementary solution to the traditional cryptography based encryption applied in the upper layers. The authors in \cite{Wang} considered PHY security in an OFDMA system, with the goal of maximizing the rate of best-effort users subject to the secure data rate requirements of confidential users. In \cite{Renna}, OFDM based wiretap channel was considered and the achievable secrecy rate with Gaussian inputs was studied. Artificial noise (AN) is another important method for improving PHY security by degrading eavesdroppers' decoding capability \cite{Goel,Qin13}. In \cite{Goel}, in order to assist secrecy information transmission, AN is transmitted into the null space of the channels of secrecy users to interfere with the eavesdroppers. In \cite{Qin13}, the authors proposed a time-domain AN scheme by exploiting temporal degrees of freedom from the cyclic prefix in OFDM modulated signals, even with a single antenna at the transmitter.
    \begin{figure}[t]
\begin{centering}
\includegraphics[scale=1.1]{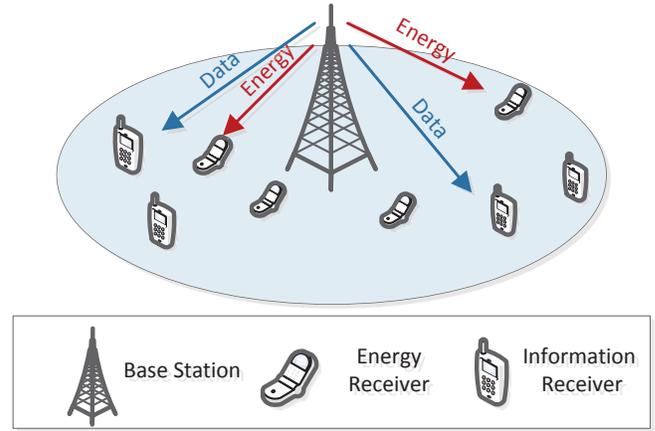}
\vspace{-0.1cm}
 \caption{System model. }\label{system:f2}
\end{centering}
\vspace{-0.3cm}
\end{figure}

\begin{figure*}[t!]
\begin{centering}
\includegraphics[scale=1]{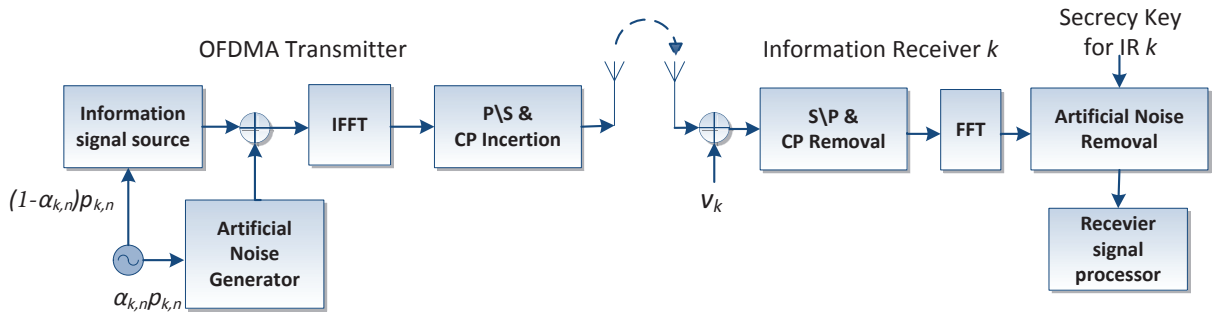}
\vspace{-0.1cm}
 \caption{Block diagram of an OFDMA transmitter with AN generation and the corresponding receiver with AN removal. }\label{fig:sys}
\end{centering}
\vspace{-0.3cm}
\end{figure*}
When PHY security is considered in SWIPT, AN not only enables secrecy information transmission but also becomes a new source for energy harvesting.
There are only a handful of works that have studied the secrecy wireless information and power transfer by properly designing the beamforming vectors at the multi-antenna transmitter \cite{LiuGC13,NgGC13}. Secrecy communication in SWIPT over fading channels was also studied in \cite{Rui14}.
However, AN aided secrecy information transmission in OFDMA-based SWIPT systems has not been addressed in the literature, which motivates this work.

In this study, we consider the resource allocation problem in the AN aided secure OFDMA systems with SWIPT, consisting of two types of receivers, i.e., information receivers (IRs) and energy receivers (ERs). We first propose a new frequency-domain AN method in OFDMA-based SWIPT to facilitate both secrecy information transmission and energy transfer to IRs and ERs, respectively. 
Specifically, independent AN signals are added over orthogonal subcarriers (SCs) and only the intended IR knows the AN transmitted over each corresponding SC and is able to remove it before decoding the information. Our goal is to maximize the weighted sum secrecy rate of the IRs subject to minimum harvested power requirements of individual ERs. The formulated problem is a mixed integer programming problem and thus is non-convex. We propose an efficient algorithm based on Lagrange duality method, which solves the problem optimally when the number of SCs becomes large.

\section{System Model and Problem Formulation}
\subsection{System Model}

    We consider a downlink OFDMA-based SWIPT system with the new consideration of PHY security as shown in Fig. \ref{system:f2}. The system consists of one base station (BS) with a single antenna, $K$ single-antenna receivers and $N$ SCs. The set of receivers is denoted by $\mathcal{K}=\{1,...,K\}$, among which $K_1$ receivers are IRs given by the set $\mathcal{K}_1$ and the other $K_2$ receivers are ERs given by the set $\mathcal{K}_2$, i.e., $\mathcal{K}_1\cup \mathcal{K}_2 = \mathcal{K}$. The set of SCs is denoted as $\mathcal{N}$. Furthermore, we assume that for each IR, all other receivers (IRs or ERs) are assumed to be potential eavesdroppers. The BS is assumed to know the channel state information (CSI) of all receivers. This is practically valid since the IRs and ERs need to help the BS obtain their individual CSI for receiving required information and energy, respectively.

    We propose a frequency-domain AN generation and removal method in OFDMA systems, similar to that in \cite{Rui14} over the time domain. The scheme is illustrated in Fig. \ref{fig:sys} and described as follows. A large ensemble of random Gaussian sequences are pre-stored at the BS, and the indices of the sequences are regarded as the keys. After SC allocation to IRs, the BS first randomly picks $N$ sequences from the ensemble (each corresponds to one SC) and transmits each of their indices (keys) to the IR to whom the corresponding SC is assigned. As the random sequence (or AN) is only known to the intended IR but unknown to all the other receivers, any potential eavesdropper cannot have access to the random sequence used at each SC. Moreover, in order to prevent the eavesdropper from decoding the random sequence by long-term observation of the signal, the BS randomly picks new random sequences and transmits the corresponding keys in a secret manner to the desired IRs from time to time, using e.g. the method proposed in \cite{Koo2000}.

    The transmit signal comprises of the transmitted data symbol $s_{k,n}$ from the BS to IR $k$ on SC $n$  and the AN bearing signal $z_{k,n}$ for IR $k$, $k \in \mathcal{K}_1$ and $n \in \mathcal{N}$. It is assumed that $s_{k,n}$ and $z_{k,n}$ are independent circularly symmetric complex Gaussian (CSCG) random variables with zero mean and unit variance, denoted by $s_{k,n}\sim\mathcal{CN}(0,1)$ and $z_{k,n}\sim\mathcal{CN}(0,1)$, which are also independent over $n$.

    The transmitted signal to IR $k$ at SC $n$ is given by
    \begin{equation}
    X_{k,n}=\sqrt{(1-\alpha_{k,n})p_{k,n}}s_{k,n}+\sqrt{\alpha_{k,n}p_{k,n}}z_{k,n},
    \end{equation}
     where $p_{k,n}\geq 0$ is the total power at SC $n$ and $0\leq\alpha_{k,n}\leq1$ is the split power ratio to generate artificial noise to be added at SC $n$, with SC $n$ assumed to be allocated to IR $k$. Notice that if $p_{k,n}>0$ and $\alpha_{k,n}=1$ for any SC $n$, then this SC is used only for energy transfer, i.e., there is no information sent over the SC.

     Let $h_{k,n}$ denote the complex channel coefficient from the BS to receiver $k$ at SC $n$, and $\beta_{k,n}$ denote the eavesdropper's complex channel coefficient. The downlink received signal at IR $k$ on SC $n$ and that at a potential eavesdropper who is wiretapping IR $k$ over SC $n$ are respectively given by
     \begin{align}
    &Y_{k,n}=h_{k,n}X_{k,n}+v_k ,\\
    &E_{k,n}=\beta_{k,n}X_{k,n}+e_k,
    \end{align}
    where the noise $v_k$ and $e_k$ are assumed to be independent and identically distributed (i.i.d.) as  $\mathcal{CN}(0,\sigma^2)$.  Here, we let $|\beta_{k,n}|^2=\max_{k' \in \mathcal{K}, k'\neq k}|h_{k',n}|^2$, indicating that the considered eavesdropper of receiver $k$
    is the receiver of the largest channel gain among all the other receivers on SC $n$. We assume that the OFDM symbols are time slotted so that the length of each time slot is comparable to the channel coherence time, i.e., the channel impulse response can be treated as time invariant during each time slot. As a result, the BS can accurately estimate $h_{k,n}$ of all receivers and $\beta_{k,n}$ on all SCs.

    With the aforementioned scheme, the AN can be removed at the desired IR at each SC but not possibly at any of the potential eavesdroppers. From (1)-(3), the received signals at IR $k$ after AN cancelation and the ``best" eavesdropper on SC $n$ are further expressed as
     \begin{align}
    &Y_{k,n}=h_{k,n}\sqrt{(1-\alpha_{k,n})p_{k,n}}s_{k,n}+v_k ,\\
    &E_{k,n}=\beta_{k,n}\sqrt{(1-\alpha_{k,n})p_{k,n}}s_{k,n}+\beta_{k,n}\sqrt{\alpha_{k,n}p_{k,n}}z_{k,n}+e_k.
    \end{align}
    Here we can write the achievable information rate of IR $k$ on SC $n$, which is given by
      \begin{align}
r_{k,n}=\log_2 \left(1+\frac{(1-\alpha_{k,n})|h_{k,n}|^2p_{k,n}}{\sigma^2}\right).
      \end{align}
The decodable rate of the ``best" eavesdropper on SC $n$ is given by
 \begin{align}
r_{k,n}^e=\log_2 \left(1+\frac{(1-\alpha_{k,n})|\beta_{k,n}|^2 p_{k,n}}{\sigma^2+\alpha_{k,n}|\beta_{k,n}|^2 p_{k,n}}\right).
 \end{align}
The achievable secrecy rate for IR $k$ on SC $n$ is thus given by \cite{Shannon}
\begin{align}\label{r}
R_{k,n}^s=&[r_{k,n}-r_{k,n}^e]^+\nonumber\\
&\left[\log_2 \left(1+\frac{(1-\alpha_{k,n})|h_{k,n}|^2p_{k,n}}{\sigma^2}\right)\right.\nonumber\\
&\left.-\log_2 \left(1+\frac{(1-\alpha_{k,n})|\beta_{k,n}|^2p_{k,n}}{\alpha_{k,n}|\beta_{k,n}|^2p_{k,n}+\sigma^2}\right)\right]^+,
\end{align}
for all $k \in \mathcal{K}_1$ and $n \in \mathcal{N}$, where $[\cdot]^+\triangleq \max(0,\cdot)$.

\begin{corollary}
$R_{k,n}^s$ in \eqref{r} can be further expressed as
\begin{equation}
R_{k,n}^s=\begin{cases}0, &~{\rm if}~ 0\leq p_{k,n}\leq [\mathcal{X}_{k,n}]^+,\\
         r_{k,n}-r_{k,n}^e, &~{\rm if}~p_{k,n}>[\mathcal{X}_{k,n}]^+, \end{cases}
\end{equation}
where
\begin{equation}
\mathcal{X}_{k,n} \triangleq \frac{\sigma^2}{\alpha_{k,n}}\left(\frac{1}{|h_{k,n}|^2}-\frac{1}{|\beta_{k,n}|^2}\right).\footnote{Note that for the case of $\alpha_{k,n}=0$, we set $[\mathcal{X}_{k,n}]^+\rightarrow+\infty$ if $|h_{k,n}|^2>|\beta_{k,n}|^2$ and $[\mathcal{X}_{k,n}]^+=0$ if $|h_{k,n}|^2\leq|\beta_{k,n}|^2$.}
\end{equation}
\end{corollary}
\begin{IEEEproof}
Please refer to Appendix A.
\end{IEEEproof}
%
%
%

The weighted sum (secrecy) rate of all $K_1$ IRs is given by
\begin{equation}\label{sr}
R_{\rm sum}^s=\sum_{k \in \mathcal{K}_1} w_k \sum_{n \in \mathcal{N}} x_{k,n} R_{k,n}^s,
\end{equation}
where $w_k$ is the positive weight of IR $k$, and $x_{k,n}$ is the binary SC allocation variable with $x_{k,n}=1$ representing SC $n$ is allocated to IR $k$ and $x_{k,n}=0$ otherwise.

Next, for the ERs, the harvested power at each ER $k \in \mathcal{K}_2$ is given by \cite{ZhouZhang}
\begin{align}
Q_{k}=\zeta_k \sum_{n \in \mathcal{N}}\left(\sum_{k \in \mathcal{K}_1} x_{k,n} p_{k,n}\right)|h_{k,n}|^2, \label{eqn:EH}
\end{align}
where $0<\zeta_{k}<1$ denotes the energy harvesting efficiency. Note that in the considered system, the ERs can harvest energy from all SCs while the IRs need orthogonal SC assignment for avoiding multiuser interference.

%
%

\subsection{Problem Formulation}
 Our goal is to maximize the weighted sum rate of the IRs by optimizing transmit power and SC allocation as well as power splitting ratio at each SC, subject to the harvested power constraints of all ERs.
 The problem can be mathematically formulated as
\begin{align}
&\max_{\bs P,\bs X,\bs \alpha} R_{\rm sum}^s \label{eqn:pb1}\\
{\rm s.t.}~~~
& Q_k \geq \bar{Q}_k, \forall k \in \mathcal{K}_2, \label{con:r}\\
&\sum_{k \in \mathcal{K}_1}\sum_{n \in \mathcal{N}} p_{k,n}x_{k,n} \leq P_{\rm max} \label{con:p1}\\
&0\leq p_{k,n} \leq P_{\rm peak}, \forall n \in \mathcal {N},k \in \mathcal{K}_1 \label{con:p2}\\
&0\leq \alpha_{k,n}\leq 1, \forall n \in \mathcal {N},k \in \mathcal{K}_1 \label{con:a}\\
&x_{k,n} \in \{0,1\}, \forall n \in \mathcal {N},k \in \mathcal{K}_1 \label{con:x1}\\
&\sum_{k \in \mathcal{K}_1} x_{k,n} \leq 1, \forall n \in \mathcal {N} ,\label{con:x2}
\end{align}
where $\bs P \triangleq \{p_{k,n}\}$ denotes the power allocation over SCs, $\bs X \triangleq \{x_{k,n}\}$ denotes the SC allocation policy, and $\bs \alpha \triangleq \{\alpha_{k,n}\}$ denotes the power splitting ratio over SCs.
In \eqref{con:r}, $\bar{Q}_k$ denotes the harvested power constraint for ER $k\in \mathcal{K}_2$. In \eqref{con:p1} and \eqref{con:p2}, $P_{\rm max}$ and $P_{\rm peak}$ represent the total power constraint over all SCs and the peak power constraint over each SC, respectively. Finally, \eqref{con:x1} and \eqref{con:x2} constrain that any SC can only be assigned to at most one IR.

\section{Proposed Solution}
The problem \eqref{eqn:pb1} is a mixed integer programming problem and thus is NP-hard. As shown in \cite{noncon}, the duality gap becomes zero in OFDM-based resource allocation problems that include the problem \eqref{eqn:pb1} as the number of SCs goes to infinity due to the time-sharing condition. This implies that problem \eqref{eqn:pb1} can be solved by the Lagrange duality method, as will be shown in this section. First, the Lagrangian function of problem \eqref{eqn:pb1} is given by
\begin{align}
&\mathcal{L}\left(\bs P,\bs \alpha, \bs X, \bs \lambda, \gamma \right)\nonumber\\
=& \sum_{k \in \mathcal{K}_1} w_k \sum_{n \in \mathcal{N}} x_{k,n} R_{k,n}^s- \gamma \left(\sum_{k \in \mathcal{K}_1}\sum_{n \in \mathcal{N}} x_{k,n}p_{k,n} - P_{\rm max}\right) \nonumber\\
&+\sum_{k \in \mathcal{K}_2}\lambda_k(Q_k-\Bar{Q}_k)\nonumber\\
=&\sum_{k \in \mathcal{K}_1} w_k\sum_{n \in \mathcal{N}} x_{k,n} R_{k,n}^s-\gamma\sum_{k \in \mathcal{K}_1}\sum_{n \in \mathcal{N}} x_{k,n}p_{k,n}\nonumber\\
&+   \sum_{n \in \mathcal{N}}\left(\sum_{k \in \mathcal{K}_1} x_{k,n} p_{k,n}\right)\sum_{k \in \mathcal{K}_2}\lambda_k\zeta_k |h_{k,n}|^2 \nonumber\\
&-\sum_{k \in \mathcal{K}_2}\lambda_k \Bar{Q}_k + \gamma P_{\rm max}\label{eqn:lf},
\end{align}
where $[\lambda_1,\lambda_2,...,\lambda_{K_2}]$ and $\gamma$ are the Lagrange multipliers (dual variables) corresponding to the minimum required harvested power constraints and the total transmit power constraint, respectively.

We then define $\mathcal{P}$ for given $\bs X$ as the set of all possible power allocations of $\bs P$ that satisfy $0\leq p_{k,n}\leq P_{\rm peak}$ for $x_{k,n}=1$ and $p_{k,n}=0$ when $x_{k,n}=0$, $\mathcal{S}$ as the set of all possible $\bs X$ that satisfy constraints \eqref{con:x1} and \eqref{con:x2}, and $\mathcal{A}$ as the set of all feasible $\bs \alpha$ that satisfy \eqref{con:a}. Then, we can obtain the dual function for the problem \eqref{eqn:pb1} as
\begin{align}\label{eqn:dual}
g(\bs \lambda, \gamma)= \max_{\bs P \in \mathcal{P}(\bs X), \bs \alpha \in \mathcal{A}, \bs X \in \mathcal{S}}\mathcal{L} \left(\bs P,\bs \alpha, \bs X, \bs \lambda, \gamma\right).
\end{align}

The dual problem is given by
\begin{align}\label{eqn:dp}
\min_{\bs \lambda\succeq0, \gamma\geq0}g(\bs \lambda, \gamma).
\end{align}

From \eqref{eqn:lf}, we can observe that the maximization in \eqref{eqn:dp} can be decomposed into $N$ independent subproblems. Hence, we can rewrite the Lagrangian as
\begin{align}
\mathcal{L}\left(\bs P,\bs \alpha, \bs X, \bs \lambda, \gamma \right)=&\sum_{n \in \mathcal{N}}\mathcal{L}_n\left(\bs P_n,\bs \alpha_n, \bs X_n\right) -\sum_{k \in \mathcal{K}_1}\lambda_k \Bar{Q}_k\nonumber\\
&+ \gamma P_{\rm max},
\end{align}
where
\begin{align}
&\mathcal{L}_n
\left(\bs P_n,\bs \alpha_n, \bs X_n\right)=w_k R_{k,n}^s - \gamma p_{k,n}
+p_{k,n}  \sum_{k \in \mathcal{K}_2}\lambda_k\zeta_k |h_{k,n}|^2.
\end{align}
And the subproblem for SC $n$ is given by
\begin{align}
&\max_{\bs P_n \in \mathcal{P}(\bs X), \bs \alpha_n \in \mathcal{A}, \bs X_n \in \mathcal{S}}\mathcal{L}_n
\left(\bs P_n,\bs \alpha_n, \bs X_n\right).
\end{align}

\subsection{Joint Optimization for Power Allocation and Power Splitting Ratio}
It is difficult to directly express the partial derivative of $R_{k,n}^s$ given in \eqref{r} with respect to either $p_{k,n}$ or $\alpha_{k,n}$. However, as we have discussed in Corollary 1, $R_{k,n}^s=0$ when $0\leq p_{k,n}\leq [\mathcal{X}_{k,n}]^+ $ and $R_{k,n}^s=r_{k,n}-r_{k,n}^e$ when $p_{k,n}>[\mathcal{X}_{k,n}]^+$. In each case, $R_{k,n}^s$ is differentiable with respect to $p_{k,n}$ and $\alpha_{k,n}$. Hence, we first find the optimal power allocation $p_{k,n}^*$ and optimal transmit power splitting ratio $\alpha_{k,n}^*$ in both cases. Then, we select $(p_{k,n}^*, \alpha_{k,n}^*)$ that achieves the largest Lagrangian.  

\subsubsection{The case of $p_{k,n}>[\mathcal{X}_{k,n}]^+$}
By using Karush-Kuhn-Tucker (KKT) conditions and combining it to the constraint \eqref{con:a}, we can obtain the optimal $\alpha_{k,n}^*$ with given $p_{k,n}$ as
\begin{align}
\alpha_{k,n}^*(p_{k,n})=\left[\frac{1}{2}+\frac{(|\beta_{k,n}|^2-|h_{k,n}|^2)\sigma^2}{2|\beta_{k,n}|^2|h_{k,n}|^2p_{k,n}}\right]_0^1\label{eqn:opta1},
\end{align}
for all $k \in \mathcal{K}_1$ and $n \in \mathcal{N}$, where $[\cdot]_{a}^{b}\triangleq\min\{\max\{\cdot,a\},b\}$.


On the other hand, by deriving the partial derivative of $\mathcal{L}_n$ with respect to $p_{k,n}$ and equating it to zero, we have
\begin{align}
a_1p_{k,n}^3+b_1p_{k,n}^2+c_1p_{k,n}+d_1=0\label{eqn:optp1},
\end{align}
where
\begin{align}
a_1=&\ln 2|h_{k,n}|^2(\alpha_{k,n}^2-\alpha_{k,n})|\beta_{k,n}|^4\Omega_{n},\\
b_1=&(\alpha_{k,n}^2-\alpha_{k,n})|\beta_{k,n}|^4|h_{k,n}|^2w_k\nonumber\\
&+\ln2|\beta_{k,n}|^2\sigma^2\left[(\alpha_{k,n}^2-1)|h_{k,n}|^2-|\beta_{k,n}|^2\alpha_{k,n}\right] \Omega_{n},\\
c_1=&\ln2(\alpha_{k,n}-1)(|h_{k,n}|^2-|\beta_{k,n}|^2)\sigma^4\Omega_{n}\nonumber\\
&+ 2(\alpha_{k,n}^2-\alpha_{k,n})|\beta_{k,n}|^2|h_{k,n}|^2w_k\sigma^2,\\
d_1=& (\alpha_{k,n}-1)(|h_{k,n}|^2-|\beta_{k,n}|^2)w_k\sigma^4-\ln2 \sigma^6\Omega_{n},\\
\Omega_{n}=&-\gamma+\sum_{k \in \mathcal{K}_2}\lambda_k\zeta_k |h_{k,n}|^2.
\end{align}
%
%
%
We define $\Phi_1(\alpha_{k,n})$ as the set of all non-negative real roots to \eqref{eqn:optp1} that satisfy $[\mathcal{X}_{k,n}]^+< p_{k,n}\leq P_{{\rm peak}}$ with given $\alpha_{k,n}$.
To jointly optimize $p_{k,n}$ and $\alpha_{k,n}$, we consider the following two subcases.

For the first subcase, we remove the $[\cdot]_0^1$ operator and assume that $\alpha_{k,n}^*(p_{k,n})$ lies in $[0,1]$. Substituting it into \eqref{eqn:optp1} to eliminate $\alpha_{k,n}$, we have
\begin{align}
a_2p_{k,n}^2+b_2p_{k,n}+c_2=0,\label{eqn:optp2}
\end{align}
where
\begin{align}
a_2=&\ln2 |\beta_{k,n}|^4|h_{k,n}|^2 \Omega_{n},\\
b_2=&w_k|\beta_{k,n}|^4 |h_{k,n}|^2+ \ln 2 \Omega_{n}|\beta_{k,n}|^2\sigma^2,\\
c_2=&\sigma^2\left\{|\beta_{k,n}|^2|h_{k,n}|^2w_k(1-|\beta_{k,n}|^2)\nonumber\right.\\
&\left.+\ln2 \Omega_{n}^2(|\beta_{k,n}|^2+|h_{k,n}|^2)\right\}.
\end{align}
Similarly, we define $\Phi_2$ as the set of all non-negative real roots to \eqref{eqn:optp2} (also the feasible candidates) that satisfy $[\mathcal{X}_{k,n}]^+< p_{k,n}\leq P_{{\rm peak}}$.

For the second subcase that $\alpha_{k,n}^*(p_{k,n})$ is greater than 1 or smaller than 0, the set of the feasible candidates for $p_{k,n}^*$ is given by $\Phi_1(\alpha_{k,n}=1)\cup\Phi_1(\alpha_{k,n}=0)$, obtained via \eqref{eqn:optp1}.
%


\subsubsection{The case of $0\leq p_{k,n}\leq[\mathcal{X}_{k,n}]^+$}
As we have discussed, $R_{k,n}^s=0$ in this case. The Lagrangian function can be thus rewritten as
\begin{align}
\mathcal{L}_n
\left(\bs P_n,\bs \alpha_n, \bs X_n\right)=
p_{k,n}  \sum_{k \in \mathcal{K}_2}\lambda_k\zeta_k |h_{k,n}|^2- \gamma p_{k,n},
\end{align}
which is a linear function with respect to $p_{k,n}$ and independent of $\alpha_{k,n}$.
The feasible candidate $p_{k,n}$ in this case can be thus obtained as
\begin{align}
p_{k,n}=\begin{cases}0, &~{\rm if}~\gamma>\sum_{k \in \mathcal{K}_2}\lambda_k\zeta_k |h_{k,n}|^2 ,\\
         \min\{[\mathcal{X}_{k,n}]^+,P_{\rm peak}\}, &~{\rm otherwise}. \end{cases}
\end{align}
Finally, combining the above discussions, we define
\begin{align}
\mathcal{F}\triangleq\Phi_1(\alpha_{k,n}=0)\cup\Phi_1(\alpha_{k,n}=1)\cup\Phi_2\cup\{0,[\mathcal{X}_{k,n}]^+,P_{\rm peak}\}\label{eqn:optp3}
\end{align}
as the set of all feasible candidates for optimal $p_{k,n}^*$. The optimal power allocation $p_{k,n}^*$ can be obtained as
\begin{align}
p_{k,n}^*= \arg \max_{p_{k,n} \in \mathcal{F}} \mathcal{L}_{n}\left(p_{k,n},\alpha_{k,n}^*(p_{k,n})\right),\label{eqn:optp4}
\end{align}
where $\alpha_{k,n}^*(p_{k,n})$ is the optimal power splitting ratio corresponding to each optimal power candidate.

%

   \begin{algorithm}[tb]
\caption{Proposed Solution for Problem \eqref{eqn:pb1}}
\begin{algorithmic}[1]
\REPEAT
\STATE Find the set of all feasible candidates $\mathcal{F}$ according to \eqref{eqn:optp1}, \eqref{eqn:optp2} and \eqref{eqn:optp3}.
\STATE Compute the corresponding $\alpha_{k,n}^*(p_{k,n})$ for all candidates.
\STATE Select the optimal $p_{k,n}^*$ and $\alpha_{k,n}^*(p_{k,n}^*)$ according to  \eqref{eqn:optp4}.
\STATE Solve SC allocation $x_{k,n}^*$ for all $k \in \mathcal{K}_1$ and $n \in \mathcal{N}$ according to \eqref{eqn:opx1}.
\STATE Update $\bs \lambda$ and $\gamma$ according to \eqref{eqn:updatel} and \eqref{eqn:updateg} respectively.
\UNTIL {$\bs \lambda$ and $\gamma$ converges.}

\end{algorithmic}

\end{algorithm}
\subsection{Subcarrier Allocation}

Substituting the optimal $\alpha_{k,n}^*$ and $p_{k,n}^*$ into $\mathcal{L}_n$, the optimal SC assignment policy is given by
   \begin{eqnarray}\label{eqn:opx1}
     x_{k,n}^{*}=\begin{cases} 1, ~{\rm if}~k=k^{*}=\arg \max_{k \in \mathcal{K}_1}\mathcal{H}_{k,n}(p_{k,n}^*,\alpha_{k,n}^*) \\
     0, ~{\rm otherwise},\end{cases}
   \end{eqnarray}
where
\begin{align}
\mathcal{H}_{k,n}(p_{k,n},\alpha_{k,n})=w_k R_{k,n}^s - \gamma p_{k,n}
+p_{k,n}  \sum_{k \in \mathcal{K}_2}\lambda_k\zeta_k |h_{k,n}|^2.
\end{align}
\subsection{Dual Update}

According to \cite{Boyd}, the dual problem is always convex; hence, the subgradient method can be used to update the dual variables to the optimal ones by an iterative procedure:
    \begin{align}
&\lambda_{k}^{t+1}=\left[\lambda_{k}^{t}-\xi_k\left(Q_k-\Bar{Q}_k\right)\right]^+, \forall k \in \mathcal{K}_2, \label{eqn:updatel}\\
&\gamma^{t+1}=\left[\gamma^{t}-\nu\left( P_{\rm max}-\sum_{n \in \mathcal{N}} \sum_{k \in \mathcal{K}_1} x_{k,n} p_{k,n} \right) \right]^+\label{eqn:updateg},
\end{align}
where $t\geq0$ is the iteration index, and $\left[\xi_1,...,\xi_{K_2}\right]$, $\nu$ are properly designed positive step-sizes.

\subsection{Complexity}
The complexity of this iterative algorithm is analyzed as follows.
For each SC, $ \mathcal{O} ({K_1}) $ computations are needed for searching the best IR. Since the optimization is independent at each SC,
the complexity is $\mathcal{O} ({K_1 N})$ for each iteration. Last, the complexity of subgradient based updates is polynomial in the number of dual variables $K_2+1$ \cite{Boyd}.
As a result, the overall complexity of the proposed algorithm for solving problem \eqref{eqn:pb1} is $\mathcal{O} ({(K_2+1)^{q}K_1 N})$, where $q$ is a positive constant.

Finally, we summarize the algorithm in Algorithm 1.

\section{Numerical Results}

   In this section, we evaluate the performance of the proposed algorithm through simulations. In the simulation setup, a single cell with radius of $200$ m is considered. The BS is located at the center of the cell. The number of the SCs is $N=64$. We assume the
   noise power of $\sigma^2=-
   83$ dBm. We consider $K_1=4$ IRs that are randomly located in the cell with distance to the BS uniformly distributed. For each IR, $w_k=1, \forall k \in \mathcal{K}_1$. We also consider $K_2=4$ ERs that are uniformly distributed with radius of $2$ m around the BS.\footnote{We consider ERs in general closer to the BS than IRs to receiver larger power (versus that of IRs used for decoding information against background noise only). However, under this circumstance, the ERs have better channel conditions than the IRs, and as a result they are more capable of eavesdropping the information \cite{LiuGC13}.} We also assume that the ERs all have the same harvested power requirement, i.e., $\Bar{Q}_k=\Bar{Q}, \forall k \in \mathcal{K}_2$.

   For performance comparison, we also consider several benchmarking schemes. First, the fixed transmit power splitting ratio with $\alpha=0.5$ or $\alpha=0.2$, is considered for complexity reduction, where the power and SC allocation is still optimized as the proposed algorithm. Here we drop the index $k,n$ in $\alpha_{k,n}$ for brevity. Second, the SC assignment is fixed (FSA) while the power allocation and transmit power splitting are jointly optimized as the proposed algorithm. Last, we consider the scheme without using AN (NoAN).

\begin{figure}[t]
\begin{centering}
\includegraphics[scale=.6]{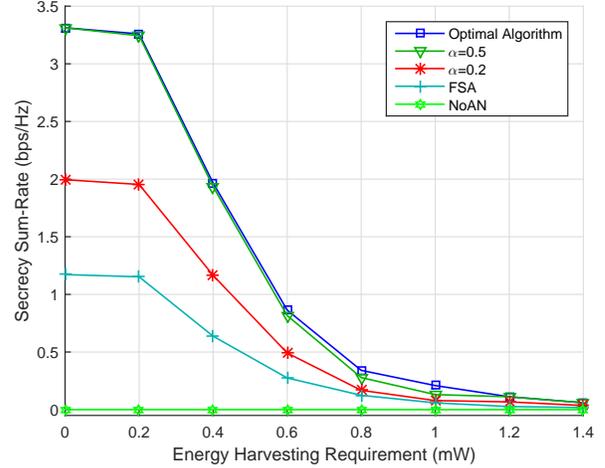}
\vspace{-0.1cm}
 \caption{Achievable secrecy sum-rate $R_{\rm sum}^s$ versus harvested power requirement. }\label{fig:f1}
\end{centering}
\vspace{-0.3cm}
\end{figure}

\begin{figure}[t]
\begin{centering}
\includegraphics[scale=.6]{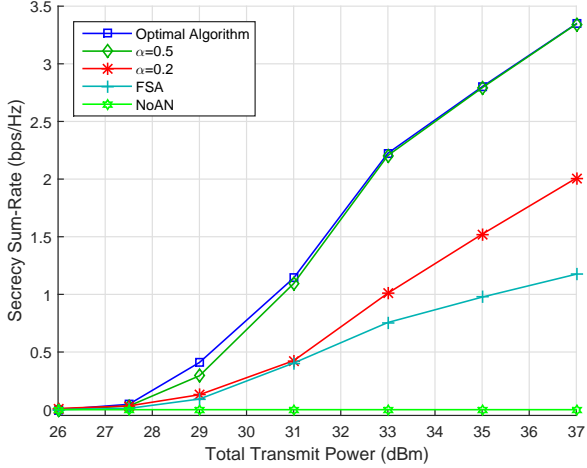}
\vspace{-0.1cm}
 \caption{Achievable secrecy sum-rate $R_{\rm sum}^s$ versus total transmit power $P_{\rm max}$. }\label{fig:f2}
\end{centering}
\vspace{-0.3cm}
\end{figure}


  In Fig. \ref{fig:f1}, the sum rate $R_{\rm sum}^s$ versus harvested power requirement $\Bar{Q}$ is demonstrated with $P_{\rm max}=37$ dBm. First, for all schemes (except NoAN), the sum rate decreases with increasing $\Bar{Q}$ (except NoAN).  An interesting observation is that the scheme with $\alpha=0.5$ performs closely to the proposed algorithm. Moreover, the proposed schemes with AN achieve great rate-energy gains over that of NoAN, which has almost zero secrecy rate regardless of harvested power requirement. This is because  without AN, the secrecy rate on each SC is positive only when it is assigned to the receiver of largest channel gain\cite{Wang}. In our simulation setup, the ERs possess much better channel gains compared to the IRs, due to much shorter distances to the BS. As a result, $|h_{k,n}|^2<|\beta_{k,n}|^2$ is true for all $n \in \mathcal{N}, k \in \mathcal{K}_1$ in general, and hence no secrecy information can be transmitted at all. This demonstrates the effectiveness of the proposed AN based approach.

  Fig. \ref{fig:f2} demonstrates the sum rate $R_{\rm sum}^s$ versus the total transmit power $P_{\rm max}$, with the harvested power requirements sets as $\Bar{Q}=100$ $\mu$W. Compared with the other benchmarking schemes, the proposed algorithm is observed to perform the best.
\section{Conclusion}

    This paper studies the optimal resource allocation problem in OFDMA-based SWIPT with the new consideration of PHY security. With a proposed frequency-domain AN generation and removal method, we maximize the weighted sum rate for the secrecy IRs subject to individual harvested power constraints of ERs by jointly optimizing transmit power and SC allocation as well as transmit power splitting ratios over SCs. The formulated problem is solved efficiently by a proposed algorithm based on Lagrange duality. Through extensive simulations, we show that the proposed algorithm outperforms other heuristically designed schemes with or without using the AN.

\appendices
\section{Proof of Corollary 1}
Equating $r_{k,n}-r_{k,n}^e$ to zero, we have
\begin{align}
\frac{|h_{k,n}|^2p_{k,n}}{\sigma^2}&=\frac{|\beta_{k,n}|^2p_{k,n}}{\alpha_{k,n}|\beta_{k,n}|^2p_{k,n}+\sigma^2}.
\end{align}

The roots of the above equation are given by $p_{k,n}=0$ and $p_{k,n}=\frac{\sigma^2}{\alpha_{k,n}}(\frac{1}{|h_{k,n}|^2}-\frac{1}{|\beta_{k,n}|^2})=\mathcal{X}_{k,n}$. However, since $p_{k,n}$ is always non-negative, $p_{k,n}=\mathcal{X}_{k,n}>0$ can be true only when $|\beta_{k,n}|^2>|h_{k,n}|^2$. Thus, we show that $r_{k,n}-r_{k,n}^e=0$ has one root at $p_{k,n}=0$, when $|h_{k,n}|^2\geq|\beta_{k,n}|^2$, and two roots at $p_{k,n}=0$ and $p_{k,n}=\mathcal{X}_{k,n}$, when $|\beta_{k,n}|^2>|h_{k,n}|^2$.

Furthermore, when $|\beta_{k,n}|^2>|h_{k,n}|^2$, it follows that
\begin{align}
&\frac{\partial(r_{k,n}-r_{k,n}^e)}{\partial p_{k,n}}|_{p_{k,n}=\mathcal{X}_{k,n}}\nonumber\\
                  =&\frac{\alpha_{k,n}(|\beta_{k,n}|^2\mathcal{X}_{k,n}^2+\mathcal{X}_{k,n}\sigma^2)}{\left(\sigma^2+\alpha_{k,n}\frac{\mathcal{X}_{k,n}}{|h_{k,n}|^2}\right)\left(\sigma^2+\frac{\mathcal{X}_{k,n}}{|h_{k,n}|^2}\right)\left[\frac{\sigma^2}{(1-\alpha_{k,n})}+\frac{\mathcal{X}_{k,n}}{|\beta_{k,n}|^2}\right]}\nonumber\\
                  \geq& 0.
\end{align}

Hence, $r_{k,n}-r_{k,n}^e\leq0$ when $0\leq p_{k,n}\leq \mathcal{X}_{k,n}$ and $|\beta_{k,n}|^2>|h_{k,n}|^2$, which is equivalent to $0\leq p_{k,n}\leq [\mathcal{X}_{k,n}]^+$. $r_{k,n}-r_{k,n}^e>0$ when $p_{k,n}> \mathcal{X}_{k,n}$, if $|\beta_{k,n}|^2>|h_{k,n}|^2$ or $p_{k,n}> 0$ if $|h_{k,n}|^2\geq |\beta_{k,n}|^2$, which is equivalent to $p_{k,n}>[\mathcal{X}_{k,n}]^+$. We finally show that $R_{k,n}^s=0$ when $0\leq p_{k,n}\leq [\mathcal{X}_{k,n}]^+$, while $R_{k,n}^s=r_{k,n}-r_{k,n}^e>0$ when $p_{k,n}>[\mathcal{X}_{k,n}]^+$.

The proof is thus completed.

\bibliographystyle{IEEEtran}
\bibliography{IEEEabrv,OFDMA}

\end{document}